# A Novel Robust Extended Dissipativity State Feedback Control system design for Interval Type-2 Fuzzy Takagi-Sugeno Large-Scale Systems

Mojtaba Asadi Jokar, Iman Zamani, Mohamad Manthouri*, Mohammad Sarbaz

Electrical and Electronic Engineering Department, Shahed University, Tehran, Iran
*Corresponding author. Tel.: +98 21 51212029.
*E-mail addresses:* asadi.jokar.mojtaba@gmail.com, zamaniiman@shahed.ac.ir, mmanthouri@shahed.ac.ir, mohammad.sarbaz@shahed.ac.ir

**Abstract:** In this paper, we use the advantage of large-scale systems modeling based on the type-2 fuzzy Takagi Sugeno model to cover the uncertainties caused by large-scale systems modeling. The advantage of using membership function information is the reduction of conservatism resulting from stability analysis. Also, this paper uses the extended dissipativity robust control performance index to reduce the effect of external perturbations on the large-scale system, which is a generalization of $H_\infty$, $L_2 - L_\infty$, passive, and dissipativity performance indexes and control gains can be achieved through solving linear matrix inequalities (LMIs), so the whole closed-loop fuzzy large-scale system is asymptotically stable.. Finally, the effectiveness of the proposed method is demonstrated by a numerical example of a double inverted pendulum system.

**Keywords:** membership function dependent, type-2 fuzzy, large-scale system, extended dissipativity, state feedback.

## I. Introduction

Since years ago, with the expansion and complexity of industries, a challenge in the engineering has been control of processes like mechanical engineering, electrical engineering, chemical engineering, or so. Many algorithms and approaches have proposed in dealing with the instability of systems. But since decades ago, systems have been turned to large in scope and dynamic. Consequently, control of the process has become an essential task and the main problem facing such systems is the complexity of mathematical relationships that make it hard to solve in practice. Various controllers have been designed to encounter the instability of systems in both industry and academic like adaptive control, fuzzy control, etc. To design an effective and authentic controller, the dynamic of the system should be identified exactly and it is an important step. In large-scale systems, it is almost impossible to identify the dynamic of the system accurately. Hence, the well-known method, fuzzy logic is used. Also, it is confirmed that the Takagi-Sugeno (T-S) model is a powerful tool to approximate any nonlinear systems with arbitrarily high accuracy. The Takagi-Sugeno fuzzy system shows the dynamic behavior of the nonlinear system with the weighted sum of local linear systems, which are determined by the membership functions.

Todays, many systems in both industry and academic are large-scale, and have been popular for years [1] and [2]. Besides, many systems specially control systems have become large in scope and complex in computation [3]. These type of the systems are modeled by a number of independent subsystems which work independently with some interactions [4]. Since years ago, lots of researches have been done in large-scale systems and gain many attentions [5], [6], and [7]. Recently, fuzzy systems with **IF-THEN** rules have become more broad appeal and most of the nonlinear and complex systems are estimated by fuzzy logic [8]. One of the powerful tools, which can fill the gap between linear and severe nonlinear systems is Takagi-Sugeno fuzzy model. Many investigations have been done based on the T-S fuzzy model [9] and [10]. [11] and [12] propose a type of fuzzy inference system popular Takagi-Sugeno, fuzzy model. In [13], Takagi-Sugeno fuzzy model is selected to represent the dynamic of the unknown nonlinear system. In [14], the stability of the fuzzy time-varying fuzzy large-scale system based on the piecewise continuous Lyapunov function investigate. Since systems always have uncertainty or perturbation parameters, several papers, including [15] and [16] have studied the robust control criterion such as $H_\infty$ control criterion for large-scale systems. [17] and [18] design a nonlinear state feedback controller for each subsystem. In [15], the stability and stabilization conditions of large-scale fuzzy systems obtained through Lyapunov functions and also by using the continuous piecewise Lyapunov functions, they have performed stability analysis and $H_\infty$ based controller design for large-scale fuzzy systems. The author in [19] design an approach for stability analysis of T-S fuzzy systems via piecewise quadratic stability. The design of robust fuzzy control for exposure to nonlinear time-delay system modeling error also presented in [20]. Also, in [16], the reference tracking problem using decentralized fuzzy $H_\infty$ control is investigated. [5] Designed decentralized linear controllers to stabilize the large-scale system using the Riccati equation basis, which increases the local state feedback gain if the number of subsystems is large.

Obviously it seems that taking the advantage of modeling systems based on type-2 fuzzy Takagi-Sugeno model to cover the uncertainties caused by large-scale systems modeling has not been studied yet and several problems remind unsolved. So the contribution of this paper can be summarized as: In the first step, we analyze the stability of the large-scale system by using a fuzzy type-2 model based on membership functions, and by using the fuzzy model type-2 decentralized state feedback controller, we stabilize the large-scale system. The advantage of using membership functions in the stability analysis is the reduction of conservatism in the stability analysis. Also, under the imperfect premise matching, the type-2 fuzzy controller can choose the premise membership functions and the number of rules different from the type-2 fuzzy model freely. Next, to reduce the effect of external perturbations on the large-scale system, we apply the robustness control criterion to the stability analysis, which can guarantee the extended dissipativity index that describes in definition1 in the appendix. Finally, a large-scale system was given to illustrate the effectiveness of the proposed method.

The rest of this paper is: Section II formulates the problem. In section III, stability conditions are given. In section IV, the numerical example presented, and section V, demonstrate the effectiveness of the proposed methods.

## II.     Problem Formulation

Consider a large-scale nonlinear system with uncertainty parameters that have $N$ subsystem and in a closed-loop system with a state feedback controller. The mathematical representation of this closed-loop system is based on the Takagi-Sugeno type-2 fuzzy model. Equation (1) shows a $p$-rule of the Takagi-Sugeno type-2 fuzzy model for the $i$th subsystem in the large-scale system:

### *Plant of Sub − System i*:

**IF** $\varsigma_{i1}(t)$ is $F_{i1}$, $\varsigma_{i2}(t)$ is $F_{i2}^l$ and … and $\varsigma_{i\psi}(t)$ is $F_{i\psi}^l$

$$\textbf{THEN} \left\{ \dot{x}_i(t) = \sum_{l=1}^{r} \widetilde{w}_{il}(x_i(t)) \left( (A_{il}x_i(t) + B_{il}u_i(t)) + D_{1il}\omega_i(t) + \sum_{\substack{k=1 \\ k \neq i}}^{N} \bar{A}_{ikl}x_k(t) \right) \right. \tag{1}$$

where $F_{i\phi}^l (\phi = 1.2 \ldots \psi)$ is a fuzzy set, and $\varsigma_{i\phi}(t)$ is a measurable variable. $r$ is the number of rules in the subsystem $i$th. $x_i(t) \in \mathbf{R}^n$ is the state vector of the $i$th subsystem. The pairs $A_{il} \in \mathbf{R}^{n \times n}, B_{il} \in \mathbf{R}^{n \times m}$ and $D_{1il}$ are matrices of the $l$th model of the $i$th subsystem. $u(t) \in \mathbf{R}^m$ denotes the input vector. $\bar{A}_{ikl}$ is the vector of the interactions between the $i$th subsystem and $k$th at the $l$th rule, and $x_k(t) \in \mathbf{R}^n$ is the state vector of the $k$th subsystem. $N$ represents the total number of subsystems. $\omega_i(t) \in \mathbf{R}^m$ is the disturbance input belonging to $\mathbf{L}_2[0,\infty)$; $\widetilde{w}_{il}(x_i(t))$ is a membership function of the $l$th rule of the $i$th subsystem, which is represented by (2).

$$\widetilde{w}_{il}(x_i(t)) = \underline{\alpha}_{il}(x_i(t))\underline{w}_{il}(x_i(t)) + \bar{\alpha}_{il}(x_i(t))\bar{w}_{il}(x_i(t)) \tag{2}$$

Equation (2) is a type reduction in the type-2 fuzzy structure in which $\underline{\alpha}_{il}$ and $\bar{\alpha}_{il}$ are nonlinear functions. As the nonlinear plant is subject to parameter uncertainties $\widetilde{w}_{il}(x_i(t))$ will depend on the parameter uncertainties and thus leads to the value of $\underline{\alpha}_{il}$ and $\bar{\alpha}_{il}$ uncertain. $\underline{w}_{il}$ and $\bar{w}_{il}$ are lower membership and upper membership degrees, respectively that characterized by the LMFs and UMFs. Since $\widetilde{w}_{il}(x_i(t))$ is a type-2 membership function, it has the following properties:

$$\sum_{l=1}^{p} \widetilde{w}_{il}(x_i(t)) = 1; \quad 0 \leq \underline{\alpha}_{il}(x_i(t)) \leq 1, \quad 0 \leq \bar{\alpha}_{il}(x_i(t)) \leq 1, \quad \forall i$$

$$\underline{\alpha}_{il}(x_i(t)) + \bar{\alpha}_{il}(x_i(t)) = 1, \quad \forall i$$

$$\underline{w}_{il}(x_i(t)) = \prod_{\alpha=1}^{\psi} \underline{\mu}_{F_{i\alpha}^l}(\varsigma_{i\alpha}(x_i(t))); \quad \bar{w}_{il}(x_i(t)) = \prod_{\alpha=1}^{\psi} \bar{\mu}_{F_{i\alpha}^l}(\varsigma_{i\alpha}(x_i(t)))$$

$$\overline{\mu}_{F_{i\alpha}^l}\left(\varsigma_{i\alpha}(x_i(t))\right) > \underline{\mu}_{F_{i\alpha}^l}\left(\varsigma_{i\alpha}(x_i(t))\right) \geq 0; \quad \overline{w}_{il}(x_i(t)) \geq \underline{w}_{il}(x_i(t)) \geq 0, \quad \forall i$$

$\underline{\mu}_{F_{i\alpha}^l}$ and $\overline{\mu}_{F_{i\alpha}^l}$ are the lower membership functions (LMF) and the upper membership functions (UMF), respectively. $\psi$ is the number of fuzzy sets of the $l$th model of the $i$th subsystem. Thus $\widetilde{w}_{il}(x_i(t))$ is a linear combination of $\underline{w}_{il}$ and $\overline{w}_{il}$ denoted by LMFs and UMFs.

Equation (3) is the Takagi-Sugeno type-2 fuzzy representation for state feedback controller. Unlike the PDC control method, the membership functions and the number of rules of the fuzzy system model and the controller need not be the same here. Thus, the membership functions and the number of controller rules relative to the plant model can be freely chosen. For the $i$th subsystem controller we have:

**Controler for Sub − System i:**
**IF** $g_{i1}(t)$ is $N_{i1}^j$, $g_{i2}(t)$ is $N_{i2}^j$ and ... and $g_{i\Omega}(t)$ is $N_{i\Omega}^j$

$$\textbf{THEN} \quad u_i(t) = \sum_{j=1}^{c} \widetilde{m}_{ij}(x_i(t)) G_{ij} x_i(t) \tag{3}$$

where $N_{i\beta}^j$ is the fuzzy set of $j$th rules of the $i$th subsystem, corresponding to the function $g_\beta(t)$. The state vector is $x_i(t) \in R^n$ where $c$ is the number of control rules of the $i$th subsystem. $G_{ij} \in R^{m\times n}$ is the control gain and $\widetilde{m}_{ij}(x_i(t))$ is the membership function of $j$th rules of the $i$th subsystem with these properties:

$$\widetilde{m}_{ij}(x_i(t)) = \frac{\underline{\beta}_{ij}(x_i(t))\underline{m}_{ij}(x_i(t)) + \overline{\beta}_{ij}(x_i(t))\overline{m}_{ij}(x_i(t))}{\sum_{j=1}^{c}(\underline{\beta}_{ij}(x_i(t))\underline{m}_{ij}(x_i(t)) + \overline{\beta}_{ij}(x_i(t))\overline{m}_{ij}(x_i(t)))} \tag{4}$$

where

$$\sum_{j=1}^{c} \widetilde{m}_{ij}(x_i(t)) = 1, \quad 0 \leq \underline{\beta}_{ij}(x_i(t)) \leq 1, \quad 0 \leq \overline{\beta}_{ij}(x_i(t)) \leq 1, \quad \forall j$$

$$\overline{\beta}_{ij}(x_i(t)) + \underline{\beta}_{ij}(x_i(t)) = 1, \quad \forall j$$

$$\underline{m}_{ij}(x_i(t)) = \prod_{\beta=1}^{\Omega} \underline{\mu}_{N_{i\beta}^j}\left(g_{i\beta}(x_i(t))\right), \quad \overline{m}_{ij}(x_i(t)) = \prod_{\beta=1}^{\Omega} \overline{\mu}_{N_{i\beta}^j}\left(g_{i\beta}(x_i(t))\right)$$

$$\overline{\mu}_{N_{i\beta}^j}\left(g_{i\beta}(x_i(t))\right) > \underline{\mu}_{N_{i\beta}^j}\left(g_{i\beta}(x_i(t))\right) \geq 0, \quad \overline{m}_{ij}(x_i(t)) > \underline{m}_{ij}(x_i(t)) \geq 0$$

where $\underline{m}_{ij}(x_i(t))$ and $\overline{m}_{ij}(x_i(t))$ denote the lower and the upper membership degree. $\underline{\beta}_{ij}(x_i(t))$ and $\overline{\beta}_{ij}(x_i(t))$ are two nonlinear functions. Relation (4) illustrates the part of the type reduction in type-2 fuzzy structure. $\Omega$ is the total number of fuzzy rules for $j$th controller rules of the $i$th subsystem. $\underline{\mu}_{N_{i\beta}^j}\left(g_{i\beta}(x_i(t))\right)$ and $\overline{\mu}_{N_{i\beta}^j}\left(g_{i\beta}(x_i(t))\right)$ represent LMF and UMF, respectively. Finally, the type-2 fuzzy model for $i$th sub-system will be:

$$\dot{x}_i(t) = \sum_{l=1}^{r}\sum_{j=1}^{r} \widetilde{h}_{ilj} \left\{ (A_{il} + B_{il}G_{ij})x_i(t) + D_{1il}\omega_i(t) + \sum_{\substack{k=1\\k\neq i}}^{N} \bar{A}_{ikl}x_k(t) \right\} \tag{5}$$

where $\widetilde{h}_{ilj}(x_i(t))$ from (4) and (4) equals the following relation:

$$\widetilde{h}_{ilj}(x_i(t)) = \widetilde{w}_{il}(x_i(t))\widetilde{m}_{ij}(x_i(t)) \tag{6}$$

has the following properties:

$$\sum_{l=1}^{p}\sum_{j=1}^{c} \widetilde{h}_{ilj}(x_i(t)) = 1, \quad \forall i,j,l$$

To facilitate the stability analysis of the large-scale type-2 fuzzy control system, we divide the state space $\phi$ into $q$ subspace, i.e., state-space equals $\phi = \bigcup_{k=1}^{q} \phi_k$. Also, to use the information of type-2 membership functions, LMFs and UMFs are described with uncertainty coverage space or briefly FOUs. Now consider dividing FOUs by $\tau + 1$ sub-FOU. In the zth sub-FOU, LMFs and UMFs define as follows:

$$\overline{h}_{iljz}(x_i(t)) = \sum_{k=1}^{q} \sum_{i1=1}^{2} \cdots \sum_{in=1}^{2} \prod_{r=1}^{n} v_{r i_r k z}(x_r(t)) \overline{\delta}_{ilji_1 i_2 \ldots i_n kz},$$

$$\underline{h}_{iljz}(x_i(t)) = \sum_{k=1}^{q} \sum_{i1=1}^{2} \cdots \sum_{in=1}^{2} \prod_{r=1}^{n} v_{r i_r k z}(x_r(t)) \underline{\delta}_{ilji_1 i_2 \ldots i_n kz} \quad (7)$$

with these properties:

$$0 \le \overline{\delta}_{ilji_1 i_2 \ldots i_n kz} \le \underline{\delta}_{ilji_1 i_2 \ldots i_n kz} \le 1, \quad 0 \le \overline{h}_{iljz}(x_i(t)) \le \underline{h}_{iljz}(x_i(t)) \le 1, \quad 0 \le v_{r i_r k i z}(x_r(t)) \le 1$$

$$v_{r1_r k i z}(x_r(t)) + v_{r2_r k i z}(x_r(t)) = 1$$

$$\sum_{k=1}^{q} \sum_{i1=1}^{2} \cdots \sum_{in=1}^{2} \prod_{r=1}^{n} v_{r i_r k z}(x_r(t)) = 1, \quad r,s = 1,2 \ldots, n; \ z = 1,2 \ldots, \tau+1; \ i_r = 1,2;$$

$x(t) \in \phi_k$; otherwise, $v_{r i_s k l}(x_r(t)) = 0$;

where $\underline{\delta}_{ilji_1 i_2 \ldots i_n kz}$ and $\overline{\delta}_{ilji_1 i_2 \ldots i_n kz}$ are scalar that must be specified. $v_{r i_r k i z}$ are functions that specified by the method intended to approximate membership functions. Finally, in order to show the sub-FOUs in $\tilde{h}_{ilj}(x_i(t))$ we have:

$$\tilde{h}_{ilj}(x_i(t)) = \tilde{w}_{il}(x_i(t)) \tilde{m}_{ij}(x_i(t))$$

$$= \sum_{z=1}^{\tau+1} \xi_{iljz}(x_i(t)) \left[ \underline{\gamma}_{iljz}(x_i(t)) \underline{h}_{iljz}(x_i(t)) + \overline{\gamma}_{iljz}(x_i(t)) \overline{h}_{iljz}(x_i(t)) \right] \quad (8)$$

where for the membership function $\tilde{h}_{ilj}(x_i(t))$ with $i,j,l$ at any one time, among $\tau + 1$ sub-FOU is only once $\xi_{iljz}(x_i(t)) = 1$ and the remainder are zero. $\underline{\gamma}_{iljz}(x_i(t))$ and $\overline{\gamma}_{iljz}(x_i(t))$ are two functions that have the following properties:

$$0 \le \underline{\gamma}_{iljz}(x_i(t)) \le \overline{\gamma}_{iljz}(x_i(t)) \le 1, \quad \overline{\gamma}_{iljz}(x_i(t)) + \underline{\gamma}_{iljz}(x_i(t)) = 1, \quad \forall l,j,z$$

### III. Main Result

In this section, we will obtain the stability of the closed-loop large-scale system using the type-2 Takagi-Sugeno model. In [11], the authors introduced a new performance index, referred to extended dissipativity performance index that holds $H_\infty$, $L_2$-$L_\infty$, passive, and dissipativity performance indexes. This performance indexes describe in definition 1 in the appendix. Therefore, the primary purpose of this section is to design the type-2 Takagi-Sugeno fuzzy state-feedback controller for the large-scale system such that the closed-loop system is asymptotically stable with the $H_\infty$, $L_2$-$L_\infty$, passive, and dissipativity performance indexes such that:

1- The closed-loop system with $\omega(t) = 0$ is asymptotically stable.
2- The closed-loop system holds extended dissipativity performance index.

**Theorem 1.** For given matrices $\phi, \psi_1, \psi_2$, and $\psi_3$ satisfying in assumption 1 in the appendix, the system in (5) is asymptotically stable and satisfies the extended dissipativity performance indexes, if there exist matrices $X_i = X_i^T > 0$, $K_i = K_i^T > 0$, $M_i = M_i^T \in \mathbf{R}^{n \times n}, N_{ij} \in \mathbf{R}^{m \times n}, W_{iljz} = W_{iljz}^T \in \mathbf{R}^{n \times n}$, $(i = 1,2, \ldots, N; l = 1,2, \ldots, p; j = 1,2, \ldots, c; z = 1,2, \ldots, \tau+1)$ such that the following LMIs hold:

$$W_{iljz} > 0 \quad \forall i,j,l,z \quad (9)$$

$$\Omega_{ilj} + W_{iljz} + M_i > 0 \quad \forall i,j,l,z \quad (10)$$

$$\sum_{l=1}^{p}\sum_{j=1}^{c}\left(\overline{\delta}_{ilji_1i_2\ldots i_nkz}\Omega_{ilj} - \left(\underline{\delta}_{ilji_1i_2\ldots i_nkz} - \overline{\delta}_{ilji_1i_2\ldots i_nkz}\right)W_{iljz} + \overline{\delta}_{ilji_1i_2\ldots i_nkz}M_i\right) - M_i$$
$$< 0 \qquad \forall i_1, i_2, \ldots, i_n, k, i, z \tag{11}$$

$$\Theta_{2i} = \begin{bmatrix} -K_i & \tilde{C}_{il}^T \phi_i^T \\ * & -I \end{bmatrix} < 0 \tag{12}$$

$$\Theta_{1i} = \begin{bmatrix} -X_i & X_i \\ * & K_i - 2I \end{bmatrix} < 0 \tag{13}$$

where

$$\widetilde{\Omega}_{ilj} = \begin{bmatrix} \widetilde{\Omega}_{1ilj} & \widetilde{\Omega}_{2il} \\ * & \widetilde{\Omega}_{3il} \end{bmatrix}, \quad \widetilde{\Omega}_{1ilj} = \boldsymbol{He}(A_{il}X_i + B_{il}N_{ij}) + \tau_0^{-1}(N-1)\left[\sum_{\substack{k=1 \\ k \neq i}}^{N}\left(X_i\tilde{A}_{ki}^T\tilde{A}_{ki}X_i\right)\right] + \tau_i - \tilde{C}_{il}^T\psi_{1i}\tilde{C}_{il}$$

$$\widetilde{\Omega}_{2il} = -D_{2il}^T\psi_{1i}D_{2il} - He(D_{2il}\psi_{2i}) - \psi_{3i}, \qquad \widetilde{\Omega}_{3il} = D_{1il} - \tilde{C}_{il}^T\psi_{1i}\tilde{C}_{il} - \tilde{C}_{il}\psi_{2i}$$

for all $i, l, j$; and the feedback gain define as $G_{ij} = N_{ij}X_i^{-1}$ for all $i, j$. Remember that $\boldsymbol{He}(A) = A + A^T$. Also $\tilde{A}_{ki}^T$ define as $\tilde{A}_{ki}^T \geq \left\|\sum_{l=1}^{p}\sum_{j=1}^{c}\tilde{h}_{ilj}(x_i(t))\bar{A}_{ikl}\right\|$.

Proof. Consider the quadratic Lyapunov function as follows:

$$V(t) = \sum_{i=1}^{N} x_i^T(t)P_i x_i(t), \quad 0 < P_i = P_i^T \in \boldsymbol{R}^{n \times n}, \quad \forall i \tag{14}$$

The main objective is to develop a condition guaranteeing that $V(t) > 0$ and $\dot{V}(t) < 0$ for all $x_i(t) \neq 0$, the type-2 fuzzy large-scale control system is guaranteed to be asymptotically stable, implying that $x_i(t) \to 0$ as $t \to \infty$. To ensure that $\dot{V}(t) < 0$ for all $x_i(t) \neq 0$ we have:

$$\dot{V}(t) = \sum_{i=1}^{N}\left\{\dot{x}_i^T(t)P_i x_i(t) + x_i^T(t)P_i \dot{x}_i(t)\right\}$$

$$= \sum_{i=1}^{N} 2\left\{\left(\sum_{l=1}^{p}\sum_{j=1}^{c}\tilde{h}_{ilj}\left\{(A_{il} + B_{il}G_{ij})x_i(t) + D_{1il}\omega_i(t)\right\}\right)^T P_i x_i(t)\right\}$$

$$+ \sum_{i=1}^{N} 2\left\{\sum_{l=1}^{p}\sum_{j=1}^{c}\tilde{h}_{ilj}\left\{\sum_{\substack{k=1 \\ k \neq i}}^{N}\bar{A}_{ikl}x_k(t)\right\}\right\}P_i x_i(t) \tag{15}$$

Same as [21] for interconnections terms by using Lemma1 in the appendix and noting that $\tilde{A}_{ik} \geq \left\|\sum_{l=1}^{p}\sum_{j=1}^{c}\tilde{h}_{ilj}(x_i(t))\bar{A}_{ikl}\right\|$ we have

$$\sum_{i=1}^{N}\left\{\left[\sum_{\substack{k=1 \\ k \neq i}}^{N}\tilde{A}_{ik}x_k(t)\right]^T\left[\sum_{\substack{k=1 \\ k \neq i}}^{N}\tilde{A}_{ik}x_k(t)\right]\right\} \leq \sum_{i=1}^{N}\left\{\left[\sum_{\substack{k=1 \\ k \neq i}}^{N}\tilde{A}_{ik}x_i(t)\right]^T\left[\sum_{\substack{k=1 \\ k \neq i}}^{N}\tilde{A}_{ik}x_i(t)\right]\right\}$$

$$\leq \sum_{i=1}^{N}\left\{(N-1)\left[\sum_{\substack{k=1 \\ k \neq i}}^{N}x_i^T(t)\tilde{A}_{ki}^T\tilde{A}_{ki}x_i(t)\right]\right\} \tag{16}$$

and by using Lemma2 in the appendix we have and by considering $0 < \tau_0 < \tau_{ilj}$ we have

$$\dot{V}(t)$$
$$\leq \sum_{i=1}^{N} 2\left\{\left(\sum_{l=1}^{p}\sum_{j=1}^{c}\tilde{h}_{ilj}\{(A_{il}+B_{il}G_{ij})x_i(t)+D_{1il}\omega_i(t)\}\right)^T P_i x_i(t)\right\}$$
$$+\sum_{i=1}^{N}\left\{\tau_0^{-1}(N-1)\left[\sum_{\substack{k=1\\k\neq i}}^{N} x_i^T(t)\tilde{A}_{ki}^T\tilde{A}_{ki}x_i(t)\right]\right\}+\sum_{i=1}^{N}\left(\sum_{l=1}^{p}\sum_{j=1}^{c}\tilde{h}_{ilj}\tau_i\left(x_i(t)^T P_i P_i x_i(t)\right)\right) \quad (17)$$

let $X_i = P_i^{-1}$, $g_i(t) = X_i^{-1} x_i(t)$, $N_{ij} = G_{ij} X_i$, $\tilde{C}_{il} = C_{il} X_i$, then we have

$$\dot{V}(t)$$
$$\leq \sum_{i=1}^{N}\left\{\sum_{l=1}^{p}\sum_{j=1}^{c}\tilde{h}_{ilj}\left(g_i^T(t)(X_i A_{il}^T + N_{ij}^T B_{il}^T + A_{il} X_i + B_{il} N_{ij})g_i(t)\right.\right.$$
$$\left.\left.+\tau_0^{-1}(N-1)\left[\sum_{\substack{k=1\\k\neq i}}^{N} g_i^T(t)(X_i \tilde{A}_{ki}^T \tilde{A}_{ki} X_i)g_i(t)\right]+\tau_i(g_i^T(t)g_i(t))\right)\right\} \quad (18)$$

$$z_i(t) = \sum_{l=1}^{p}\sum_{j=1}^{c}\tilde{h}_{ilj}\{\tilde{C}_{il}g_i(t)+D_{2il}\omega_i(t)\} \quad (19)$$

now by consider the following performance index we have

$$\dot{V}(t)-J(t) \leq \sum_{i=1}^{N}\zeta_i^T\left\{\sum_{l=1}^{p}\sum_{j=1}^{c}\tilde{h}_{ilj}\tilde{\Omega}_{ilj}\right\}\zeta_i \quad (20)$$

$$J(t) = \sum_{i=1}^{N}\left(z_i^T \psi_{1i} z_i + 2z_i^T \psi_{2i}\omega_i(t) + \omega_i^T(t)\psi_{3i}\omega_i(t)\right)$$
$$= \sum_{i=1}^{N}(\sum_{l=1}^{p}\sum_{j=1}^{c}\tilde{h}_{ilj}\{\tilde{C}_{il}g_i(t)+D_{2il}\omega_i(t)\}^T \psi_{1i}\{\tilde{C}_{il}g_i(t)+D_{2il}\omega_i(t)\}$$
$$+ 2\{\tilde{C}_{il}g_i(t)+D_{2il}\omega_i(t)\}^T \psi_{2i}\omega_i(t)+\omega_i^T(t)\psi_{3i}\omega_i(t) \quad (21)$$

where

$\zeta_i(t) = \begin{bmatrix} g_i(t) \\ \omega_i(t) \end{bmatrix}$, $\tilde{\Omega}_{ilj} = \begin{bmatrix} \tilde{\Omega}_{1ilj} & \tilde{\Omega}_{2il} \\ * & \tilde{\Omega}_{3il} \end{bmatrix}$, $\tilde{\Omega}_{1ilj} = He(A_{il}X_i + B_{il}N_{ij}) + \tau_0^{-1}(N-1)\left[\sum_{\substack{k=1\\k\neq i}}^{N}(X_i\tilde{A}_{ki}^T\tilde{A}_{ki}X_i)\right] + \tau_i - \tilde{C}_{il}^T\psi_{1i}\tilde{C}_{il}$, $\tilde{\Omega}_{2il} = -D_{2il}^T\psi_{1i}D_{2il} - He(D_{2il}\psi_{2i}) - \psi_{3i}$, $\tilde{\Omega}_{3il} = D_{1il} - \tilde{C}_{il}^T\psi_{1i}\tilde{C}_{il} - \tilde{C}_{il}\psi_{2i}$.

by using Schur complement we have

$\Omega_{ilj} = \begin{bmatrix} \Omega_{11ilj} & \Omega_{12il} & \Omega_{13il} \\ * & \Omega_{22il} & \Omega_{23il} \\ * & * & -I \end{bmatrix}$, $\Omega_{11ilj} = He(A_{il}X_i + B_{il}N_{ij}) + \tau_0^{-1}(N-1)\left[\sum_{\substack{k=1\\k\neq i}}^{N}(X_i\tilde{A}_{ki}^T\tilde{A}_{ki}X_i)\right] + \tau_i$, $\Omega_{12il} = D_{1il} - \tilde{C}_{il}\psi_{2i}$, $\Omega_{13il} = \tilde{C}_{il}^T\psi_{1i}^T$, $\Omega_{22il} = -He(D_{2il}\psi_{2i}) - \psi_{3i}$, $\Omega_{23il} = D_{2il}^T\psi_{1i}^T$.

if we can prove $\sum_{l=1}^{r}\sum_{j=1}^{r}\tilde{h}_{ilj}\Omega_{ilj} < 0$ then we have:

$$\dot{V}(t) - J(t) \leq \sum_{i=1}^{N} \zeta_i^T \left\{ \sum_{l=1}^{p} \sum_{j=1}^{c} \tilde{h}_{ilj} \tilde{\Omega}_{ilj} \right\} \zeta_i < 0 \qquad (22)$$

now by using (8) considering the information of the sub-FOUs is brought to the stability analysis with the introduction of some slack matrices through the following inequalities using the S-procedure:

let $M_i = M_i^T$ is an arbitrary matrix with appropriate dimensions. Then

$$\left\{ \sum_{l=1}^{p} \sum_{j=1}^{c} \sum_{z=1}^{\tau+1} \xi_{iljz}(x_i(t)) \left[ \left( \underline{\gamma}_{iljz}(x_i(t)) \underline{h}_{iljz}(x_i(t)) + \overline{\gamma}_{iljz}(x_i(t)) \overline{h}_{iljz}(x_i(t)) \right) - 1 \right] M_i \right\} = 0 \qquad (23)$$

also, consider $0 \leq W_{iljz} = W_{iljz}^T$

$$-\sum_{l=1}^{p} \sum_{j=1}^{c} \left(1 - \underline{\gamma}_{iljz}(x_i(t))\right) \left(\underline{h}_{iljz}(x_i(t)) - \overline{h}_{iljz}(x_i(t))\right) W_{iljz} \geq 0 \qquad (24)$$

by using (23) and (24) for $\sum_{l=1}^{r} \sum_{j=1}^{r} \tilde{h}_{ilj} \Omega_{ilj} < 0$ we have

$$\sum_{i=1}^{N} \left\{ \sum_{l=1}^{p} \sum_{j=1}^{c} \sum_{z=1}^{\tau+1} \left( \xi_{iljz}(x_i(t)) \left[ \underline{\gamma}_{iljz}(x_i(t)) \underline{h}_{iljz}(x_i(t)) + \left(1 - \underline{\gamma}_{iljz}(x_i(t))\right) \overline{h}_{iljz}(x_i(t)) \right] \right) \Omega_{ilj} \right\}$$

$$- \sum_{i=1}^{N} \sum_{l=1}^{p} \sum_{j=1}^{c} \sum_{z=1}^{\tau+1} \xi_{iljz}(x_i(t)) \left(1 - \underline{\gamma}_{iljz}(x_i(t))\right) \left(\underline{h}_{iljz}(x_i(t)) - \overline{h}_{iljz}(x_i(t))\right) W_{iljz}$$

$$+ \sum_{i=1}^{N} \left\{ \sum_{l=1}^{p} \sum_{j=1}^{c} \sum_{z=1}^{\tau+1} \xi_{iljz}(x_i(t)) \left[ \left( \underline{\gamma}_{iljz}(x_i(t)) \underline{h}_{iljz}(x_i(t)) + \left(1 - \underline{\gamma}_{iljz}(x_i(t))\right) \overline{h}_{iljz}(x_i(t)) \right) \right. \right.$$

$$\left. \left. - 1 \right] M_i \right\}$$

$$= \sum_{i=1}^{N} \left\{ \left[ \sum_{l=1}^{p} \sum_{j=1}^{c} \sum_{z=1}^{\tau+1} \xi_{iljz}(x_i(t)) \left( \overline{h}_{iljz}(x_i(t)) \Omega_{ilj} - \left( \underline{h}_{iljz}(x_i(t)) - \overline{h}_{iljz}(x_i(t)) \right) W_{iljz} \right. \right. \right.$$

$$\left. \left. \left. + \overline{h}_{iljz}(x_i(t)) M_i \right) - M_i \right] \right\}$$

$$+ \sum_{i=1}^{N} \sum_{l=1}^{p} \sum_{j=1}^{c} \sum_{z=1}^{\tau+1} \xi_{iljz}(x_i(t)) \underline{\gamma}_{iljz}(x_i(t)) \left(\underline{h}_{iljz}(x_i(t)) - \overline{h}_{iljz}(x_i(t))\right) \left(\Omega_{ilj} + W_{iljz} + M_i\right) < 0 \qquad (25)$$

also, the following equation must be checked

$$\left[ \sum_{l=1}^{p} \sum_{j=1}^{c} \sum_{z=1}^{\tau+1} \xi_{iljz}(x_i(t)) \left( \overline{h}_{iljz}(x_i(t)) \Omega_{ilj} - \left(\underline{h}_{iljz}(x_i(t)) - \overline{h}_{iljz}(x_i(t))\right) W_{iljz} + \overline{h}_{iljz}(x_i(t)) M_i \right) \right.$$

$$\left. - M_i \right] < 0 \quad \forall i \qquad (26)$$

and $\tilde{\Omega}_{ilj} + W_{iljz} + M_i > 0$ for all $i,j,l,z$ due to $\left(\underline{h}_{iljz}(x_i(t)) - \overline{h}_{iljz}(x_i(t))\right) \leq 0$. Recalling that only one $\xi_{iljz}(x_i(t)) = 1$ for each fixed value of $i,l,j$ at any time instant such that $\sum_{z=1}^{\tau+1} \xi_{iljz}(x_i(t)) = 1$, the first set of inequality is satisfied by

$$\left[\sum_{l=1}^{p}\sum_{j=1}^{c}\left(\overline{h}_{iljz}(x_i(t))\Omega_{ilj}-\left(\underline{h}_{iljz}(x_i(t))-\overline{h}_{iljz}(x_i(t))\right)W_{iljz}+\overline{h}_{iljz}(x_i(t))M_i\right)-M_i\right]$$
$$<0 \quad \forall i,l,j,z \tag{27}$$

Expressing $\overline{h}_{iljz}(x_i(t))$ and $\underline{h}_{iljz}(x_i(t))$ with (7) and recalling that $\sum_{k=1}^{q}\sum_{i1=1}^{2}\ldots\sum_{in=1}^{2}\prod_{r=1}^{n}v_{r_{i_r}kiz}(x_r(t))=1$, for all $z$ and $v_{r_{i_r}kiz}\geq 0$ for all $r,i_r,k,i$ and $z$ the first set of inequalities will be satisfied if the following inequalities hold

$$\left[\sum_{k=1}^{q}\sum_{i1=1}^{2}\ldots\sum_{in=1}^{2}\prod_{r=1}^{n}v_{r_{i_r}kz}(x_r(t))\sum_{l=1}^{p}\sum_{j=1}^{c}(\overline{\delta}_{ilji_1i_2\ldots i_nkz}\Omega_{ilj}-(\underline{\delta}_{ilji_1i_2\ldots i_nkz}-\overline{\delta}_{ilji_1i_2\ldots i_nkz})W_{iljz}\right.$$
$$\left.+\overline{\delta}_{ilji_1i_2\ldots i_nkz}M_i)-M_i\right]<0 \quad \forall i_1,i_2,\ldots,i_n,k,i,z \tag{28}$$

consequently (27) can be guaranteed by

$$\sum_{l=1}^{p}\sum_{j=1}^{c}(\overline{\delta}_{ilji_1i_2\ldots i_nkz}\Omega_{ilj}-(\underline{\delta}_{ilji_1i_2\ldots i_nkz}-\overline{\delta}_{ilji_1i_2\ldots i_nkz})W_{iljz}+\overline{\delta}_{ilji_1i_2\ldots i_nkz}M_i)-M_i$$
$$<0 \quad \forall i_1,i_2,\ldots,i_n,k,i,z \tag{29}$$

therefore, there is always a sufficiently small scalar $c>0$ such that $\widetilde{\Omega}_{ilj}\leq -cI$. This means that

$$\dot{V}(t)-J(t)\leq -c\left|\sum_{i=1}^{N}\zeta_i\right|^2 \tag{30}$$

thus $J(t)\geq \dot{V}(t)$ hold for any $t\geq 0$, which means

$$\int_0^t J(s)\,ds \geq V(x(t))-V(x(0)) \tag{31}$$

then by considering $\rho=-V(x(0))$ in (31) we have:

$$\int_0^t J(s)\,ds \geq V(x(t))+\rho, \quad \forall t\geq 0 \tag{32}$$

According to Definition 1 in the appendix, if we want to design a controller with a robust $H_\infty$ performance, then we must set the $\rho$ value to zero. For substitution $V(x(t))$ in (32) considering $K_i>0$ by Characteristic $(K_i-I)K_i^{-1}(K_i-I)\geq 0$ where $-K_i^{-1}\leq K_i-2I$ then we have:

$$\Theta_{1i}=\begin{bmatrix}-X_i & X_i\\ * & K_i-2I\end{bmatrix}<0 \tag{33}$$

Finally $P_i>K_i$ and (14) proved if :

$$V(x(t))=\sum_{i=1}^{N}x_i^T(t)P_ix_i(t)\geq \sum_{i=1}^{N}x_i^T(t)K_ix_i(t)\geq 0 \tag{34}$$

Also

$$V(x(t))=\sum_{i=1}^{N}x_i^T(t)P_ix_i(t)\geq \sum_{i=1}^{N}x_i^T(t)K_ix_i(t)\geq 0 \tag{35}$$

According to Definition 1, we need to prove that the following inequality holds for any matrices $\phi_i,\psi_{1i},\psi_{2i}$ and $\psi_{3i}$ satisfying assumption 1 in the appendix:

$$\int_0^t J(t)\,dt-z^T(t)\phi z(t)\geq \rho \tag{36}$$

To this end, we consider the two cases of $\|\phi\| = 0$ and $\|\phi\| \neq 0$, respectively. Firstly, we consider the case when $\|\phi\| = 0$. Also in this case by considering $\psi_1 = -I$, $\psi_2 = 0$, $\psi_3 = \gamma^2 I$ and $\rho = 0$ the $H_\infty$ performance index will be hold.

$$\int_0^t J(s)\,ds = \sum_{i=1}^N x_i^T(t) K_i x_i(t) + \rho \geq \rho, \quad \forall t \geq 0 \tag{37}$$

By using (37) and considering $z^T(t)\phi z(t) \equiv 0$ the (36) hold. Secondly, we consider the case of $\|\phi\| \neq 0$. In this case, it is required under Assumption1 in the appendix that $\|\psi_1\| + \|\psi_2\| = 0$ and $\|D_{2il}\| = 0$, which implies that $\psi_1 = 0$, $\psi_2 = 0$ and $\psi_3 > 0$. Then:

$$J(s) = \sum_{i=1}^N \omega_i^T(s) \psi_{3i} \omega_i^T(s) \geq 0$$

now by considering $\tilde{C}_{il}^T \phi_i \tilde{C}_{il} \leq K_i$ due to

$$\Theta_{2i} = \begin{bmatrix} -K_i & \tilde{C}_{il}^T \phi_i^T \\ * & -I \end{bmatrix} < 0 \tag{38}$$

and $\|D_{2il}\| = 0$ satisfy in assumption1 for any $t \geq 0$, the following inequalities hold

$$\int_0^t J(s)\,ds - z^T(t)\phi z(t)$$

$$\geq \int_0^t J(s)\,ds - \sum_{i=1}^N \sum_{l=1}^p \sum_{j=1}^c \tilde{h}_{ilj} \{C_{il} x_i(t) + D_{2il}\omega_i(t)\}^T \phi_i \{C_{il} x_i(t) + D_{2il}\omega_i(t)\}$$

$$= \int_0^t J(s)\,ds - \sum_{i=1}^N \sum_{l=1}^p \sum_{j=1}^c \tilde{h}_{ilj} \left( g_i^T(t) \tilde{C}_{il}^T \phi_i \tilde{C}_{il} g_i(t) \right)$$

$$\geq \int_0^t J(s)\,ds - \sum_{i=1}^N \sum_{l=1}^p \sum_{j=1}^c \tilde{h}_{ilj} \left( x_i^T(t) K_i x_i(t) \right) \geq \rho \tag{39}$$

Finally by $\omega(t) \equiv 0$ we have:

$$\dot{V}(t) \leq z^T(t) \left( \sum_{i=1}^N \sum_{l=1}^p \sum_{j=1}^c \tilde{h}_{ilj}(\psi_{1il}) \right) z(t) - c \left| \sum_{i=1}^N \zeta_i \right|^2 \tag{40}$$

According to assumption1 in the appendix $\psi_{1il} < 0$ for any $i, l$, then we have

$$\dot{V}(t) \leq -c \left| \sum_{i=1}^N \zeta_i \right|^2 \tag{41}$$

thus the closed-loop system asymptotically stable by $\omega(t) \equiv 0$. This completes the proof.

Remark1: It can be seen from (8) that if more sub-FOUs are considered the more information about the FOU is contained in the local LMFs and UMFs. Thus, using the information of membership functions into the stability condition is resulting in a more relaxed stability analysis result.

Remark2: From (28), the advantage of using the type-2 fuzzy system in the form of (5) can be seen that local LMFs and UMFs determine the stability condition.

Remark3: By expressing $\overline{h}_{iljz}(x_i(t))$ and $\underline{h}_{iljz}(x_i(t))$ in the form of (7), they are characterized by the constant scalers $\overline{\delta}_{ilji_1 i_2 \ldots i_n kz}$ and $\underline{\delta}_{ilji_1 i_2 \ldots i_n kz}$. Also, noting that the cross terms $\prod_{r=1}^n v_{r i_r kz}(x_r(t))$ are independent of $i$ and $l$. By these favorable properties we need only to check (28) at some discrete points ($\overline{\delta}_{ilji_1 i_2 \ldots i_n kz}$ and $\underline{\delta}_{ilji_1 i_2 \ldots i_n kz}$) instead of every single point of the local LMFs and UMFs.

Remark4: Under the imperfect premise matching, the type-2 fuzzy controller can choose the premise membership functions and the number of rules different from the type-2 fuzzy model freely.

**Corollary:** In the particular case, if we do not consider disturbance, then we have the following result. First, we consider a large-scale nonlinear system that is composed of $N$ nonlinear subsystems with interconnections. A p-rule type-2 fuzzy T-S model is employed to describe the dynamics of the $i$th nonlinear subsystem as follows:

**Plant Rule $l$:**
**IF** $\varsigma_{i1}(t)$ is $F_{i1}^l$, $\varsigma_{i2}(t)$ is $F_{i2}^l$ and ... and $\varsigma_{i\psi}(t)$ is $F_{i\psi}^l$
**THEN**

$$\dot{x}_i(t) = \sum_{l=1}^{r} \widetilde{w}_{il}(x_i(t)) \left( (A_{il}x_i(t) + B_{il}u_i(t)) + \sum_{\substack{k=1 \\ k \neq i}}^{N} \bar{A}_{ikl}x_k(t) \right) \tag{42}$$

where $F_{i\alpha}^l$ is a type-2 fuzzy set of rule $l$ corresponding to the function $\varsigma_{i\alpha}(t)$, $i = 1,2,\ldots,N$; $\alpha = 1,2,\ldots,\psi$; $l = 1,2,\ldots,p$; $\psi$ is a positive integer; $x_i(t) \in \mathbf{R}^n$ is the $i$th subsystem state vector; the $A_{il} \in \mathbf{R}^{n \times n}$ and $B_{il} \in \mathbf{R}^{n \times m}$ are the known system and input matrices, respectively; $u_i \in \mathbf{R}^m$ is the input vector. $\bar{A}_{ikl}$ denotes the interconnection matrix between the $i$th and $k$th subsystems; $(A_{il}, B_{il})$ are the $l$th local model; The firing strength of the $p$th rule of $i$th subsystem is of the form (2). Like controller in (3) the membership functions and the number of rules of the fuzzy system model and the controller need not be the same here. Thus, the membership functions and the number of controller rules relative to the plant model can be freely chosen. For the $i$th subsystem controller we have:

**Controller Rule $l$:**
**IF** $g_1(x(t))$ is $N_{i1}^j$, $g_{i2}(x(t))$ is $N_{i2}^j$ and ... and $g_{i\Omega}(x(t))$ is $N_{i\Omega}^j$

$$\textbf{THEN } u_i(t) = \sum_{j=1}^{r} \widetilde{m}_{ij}(x_i(t)) G_{ij}x_i(t) \tag{43}$$

where $N_{i\beta}^j$ is a type-2 fuzzy set of rule $j$th corresponding to the function $g_{i\beta}(x(t))$, $\beta = 1,2,\ldots,\Omega$; $j = 1,2,\ldots,c$; $\Omega$ is a positive integer; $G_j \in \mathbf{R}^{m \times n}$ are the constant feedback gains to be determined. The firing strength of the $j$th rule is the form of (4). Finally, we have the following type-2 fuzzy T-S large-scale control system:

$$\dot{x}_i(t) = \sum_{l=1}^{r} \sum_{j=1}^{r} \widetilde{w}_{il} \widetilde{m}_{ij} \left\{ (A_{il} + B_{il}G_j)x_i(t) + \sum_{\substack{k=1 \\ k \neq i}}^{N} \bar{A}_{ikl}x_k(t) \right\} \tag{44}$$

Now, decentralized state feedback type-2 fuzzy T-S controller design presented for the continuous-time large-scale type-2 fuzzy T-S model system in (55).

**Theorem 2.** Consider a large-scale type-2 fuzzy T-S system model in (42). Decentralized state feedback type-2 fuzzy controller in the form of (43) exist, and can guarantee the asymptotic stability of the closed-loop type-2 fuzzy control system (44) if there exist $X_i = X_i^T > 0$, $G_i = G_i^T > 0$, $M_i = M_i^T \in \mathbf{R}^{n \times n}$, $N_{ij} \in \mathbf{R}^{m \times n}$, $W_{iljz} = W_{iljz}^T \in \mathbf{R}^{n \times n}$, ($i = 1,2,\ldots,N$; $l = 1,2,\ldots,p$; $j = 1,2,\ldots,c$; $z = 1,2,\ldots,\tau+1$) such that the following LMIs hold:

$$W_{iljz} > 0 \quad \forall i,j,l,z \tag{45}$$

$$\left( \left( (X_i A_{il}^T + N_{ij}^T B_{il}^T + A_{il}X_i + B_{il}N_{ij}) + \tau_0^{-1}(N-1) \left[ \sum_{\substack{k=1 \\ k \neq i}}^{N} (X_i \tilde{A}_{ki}^T \tilde{A}_{ki} X_i) \right] + \tau_i I \right) + W_{iljz} + M_i \right)$$
$$> 0 \quad \forall i,j,l,z \tag{46}$$

$$\sum_{l=1}^{p}\sum_{j=1}^{c}\left(\overline{\delta}_{ilji_1i_2...i_nkz}\left((X_iA_{il}^T+N_{ij}^TB_{il}^T+A_{il}X_i+B_{il}N_{ij})+\tau_0^{-1}(N-1)\left[\sum_{\substack{k=1\\k\neq i}}^{N}(X_i\tilde{A}_{ki}^T\tilde{A}_{ki}X_i)\right]\right.\right.$$
$$\left.\left.+\tau_iI\right)-(\underline{\delta}_{ilji_1i_2...i_nkz}-\overline{\delta}_{ilji_1i_2...i_nkz})W_{iljz}+\overline{\delta}_{ilji_1i_2...i_nkz}M_i\right)-M_i<0 \quad \forall i_1,i_2,...,i_n,k,i,z$$

(47)

where $\overline{\delta}_{ilji_1i_2...i_nkz}$ and $\underline{\delta}_{ilji_1i_2...i_nkz}$, $i=1,2,...,N; l=1,2,...,p; j=1,2,...,c; z=1,2,...,\tau+1; i_n=1,2; k=1,2,...,q$ are predefine constant scalers satisfying (7).

**Proof.** We consider the following quadratic Lyapunov function candidate to investigate the stability of the type-2 fuzzy T-S large-scale control system

$$V(t)=\sum_{i=1}^{N}x_i^T(t)P_ix_i(t) \quad (48)$$

where $0<P_i=P_i^T\in \mathbf{R}^{n\times n}$.

The main objective is to develop a condition guaranteeing that $V(t)>0$ and $\dot{V}(t)<0$ for all $x_i(t)\neq 0$, the type-2 fuzzy T-S large-scale control system is guaranteed to be asymptotically stable, implying that $x_i(t)\to 0$ as $t\to\infty$. We have:

$$\dot{V}(t)=\sum_{i=1}^{N}\{\dot{x}_i^T(t)P_ix_i(t)+x_i^T(t)P_i\dot{x}_i(t)\}$$
$$=\sum_{i=1}^{N}\left\{\left(\sum_{l=1}^{p}\sum_{j=1}^{c}\tilde{w}_{il}\tilde{m}_{ij}\left\{(A_{il}+B_{il}G_{ij})x_i(t)+\sum_{\substack{k=1\\k\neq i}}^{N}\bar{A}_{ik}x_k(t)\right\}\right)^T P_ix_i(t)\right.$$
$$\left.+x_i^T(t)P_i\left(\sum_{l=1}^{p}\sum_{j=1}^{c}\tilde{w}_{il}\tilde{m}_{ij}\left\{(A_{il}+B_{il}G_{ij})x_i(t)+\sum_{\substack{k=1\\k\neq i}}^{N}\bar{A}_{ik}x_k(t)\right\}\right)\right\} \quad (49)$$

Such as proof of Theorem1 for interconnection term by considering $\tilde{A}_{ki}\geq\left\|\sum_{l=1}^{p}\sum_{j=1}^{c}\tilde{h}_{ilj}\bar{A}_{ikl}\right\|$. we have:

$$\dot{V}(t)$$
$$\leq\sum_{i=1}^{N}2\left\{\left(\sum_{l=1}^{p}\sum_{j=1}^{c}\tilde{h}_{ilj}\{(A_{il}+B_{il}G_{ij})x_i(t)\}\right)^T P_ix_i(t)\right\}$$
$$+\sum_{i=1}^{N}\left\{\tau_0^{-1}(N-1)\left[\sum_{\substack{k=1\\k\neq i}}^{N}x_i^T(t)\tilde{A}_{ki}^T\tilde{A}_{ki}x_i(t)\right]\right\}+\sum_{i=1}^{N}\left(\sum_{l=1}^{p}\sum_{j=1}^{c}\tilde{h}_{ilj}\tau_i\left(x_i(t)^TP_iP_ix_i(t)\right)\right) \quad (50)$$

let $X_i=P_i^{-1}, \phi_i(t)=X_i^{-1}x_i(t), N_{ij}=G_{ij}X_i$, then we have

$$\dot{V}(t)$$
$$= \sum_{i=1}^{N} \left\{ \sum_{l=1}^{p} \sum_{j=1}^{c} \tilde{h}_{ilj}(x_i(t)) \left( \phi_i^T(t)(X_i A_{il}^T + N_{ij}^T B_{il}^T + A_{il} X_i + B_{il} N_{ij}) \phi_i(t) \right. \right.$$
$$\left. \left. + \tau_0^{-1}(N-1) \left[ \sum_{\substack{k=1 \\ k \neq i}}^{N} \phi_i^T(t)(X_i \tilde{A}_{ki}^T \tilde{A}_{ki} X_i) \phi_i(t) \right] + \tau_i(\phi_i^T(t)\phi_i(t)) \right) \right\}$$
(51)

we then express the type-2 membership function in the form of (8) and by considering the information of the sub-FOUs brought to stability analysis with the introduction of some slack matrices as in (23) and (24). Then we have $\dot{V}(t) < 0$ for all $x_i(t) \neq 0$ from:

$$\left[ \sum_{l=1}^{p} \sum_{j=1}^{c} \sum_{z=1}^{\tau+1} \xi_{iljz}(x_i(t)) \left( \overline{h}_{iljz}(x_i(t)) \left( (X_i A_{il}^T + N_{ij}^T B_{il}^T + A_{il} X_i + B_{il} N_{ij}) \right. \right. \right.$$
$$\left. \left. + \tau_0^{-1}(N-1) \left[ \sum_{\substack{k=1 \\ k \neq i}}^{N} \left( X_i \bar{A}_{kilj}^T \bar{A}_{kilj} X_i \right) \right] + \tau_{ilj} I \right) - \left( \underline{h}_{iljz}(x_i(t)) - \overline{h}_{iljz}(x_i(t)) \right) W_{iljz} \right.$$
$$\left. \left. + \overline{h}_{iljz}(x_i(t)) M_i \right) - M_i \right] < 0 \quad \forall i$$
(52)

(52) satisfied if the following inequality hold:

$$\left[ \sum_{l=1}^{p} \sum_{j=1}^{c} \left( \overline{h}_{iljz}(x_i(t)) \left( (X_i A_{il}^T + N_{ij}^T B_{il}^T + A_{il} X_i + B_{il} N_{ij}) + \tau_0^{-1}(N-1) \left[ \sum_{\substack{k=1 \\ k \neq i}}^{N} \left( X_i \bar{A}_{kilj}^T \bar{A}_{kilj} X_i \right) \right] \right. \right. \right.$$
$$\left. \left. + \tau_{ilj} I \right) - \left( \underline{h}_{iljz}(x_i(t)) - \overline{h}_{iljz}(x_i(t)) \right) W_{iljz} + \overline{h}_{iljz}(x_i(t)) M_i \right) - M_i \right] < 0 \quad \forall i, l, j, z$$
(53)

also, the second set of inequalities will be satisfied if the following inequalities hold:

$$\sum_{l=1}^{p} \sum_{j=1}^{c} \left( \overline{\delta}_{ilji_1 i_2 \ldots i_n kz} \left( (X_i A_{il}^T + N_{ij}^T B_{il}^T + A_{il} X_i + B_{il} N_{ij}) + \tau_0^{-1}(N-1) \left[ \sum_{\substack{k=1 \\ k \neq i}}^{N} \left( X_i \bar{A}_{kilj}^T \bar{A}_{kilj} X_i \right) \right] \right. \right.$$
$$\left. \left. + \tau_{ilj} I \right) - (\underline{\delta}_{ilji_1 i_2 \ldots i_n kz} - \overline{\delta}_{ilji_1 i_2 \ldots i_n kz}) W_{iljz} + \overline{\delta}_{ilji_1 i_2 \ldots i_n kz} M_i \right) - M_i < 0 \quad \forall i_1, i_2, \ldots, i_n, k, i, z$$
(54)

This completes the proof.

**IV.     Simulation**

Consider a double-inverted pendulum system connected by a spring, the modified equations of the motion for the interconnected pendulum are given by [21].

$$\begin{cases} \dot{x}_{i1} = x_{i2} \\ \dot{x}_{i2} = -\dfrac{kr^2}{4J_i}x_{i1} + \dfrac{kr^2}{4J_i}\sin(x_{i1})x_{i2} + \dfrac{2}{J_i}x_{i2} + \dfrac{1}{J_i}u_i + \sum_{\substack{j=1 \\ j\neq i}}^{2}\dfrac{kr^2}{8J_i}x_{j1} \;, i = \{1,2\} \end{cases} \quad (55)$$

where $x_{i1}$ denotes the angle of the $i$th pendulum from the vertical; $x_{i2}$ is the angular velocity of the $i$th pendulum. The objective here is to design robust decentralized state feedback $H_\infty$ fuzzy type-2 controller for the T-S fuzzy type-2 large-scale in the form of such that the resulting closed-loop system is asymptotically stable with an $H\infty$ disturbance attenuation level $\gamma$. A concise framework on the decentralized state feedback control shown in Fig. 1. A concise framework on the decentralized State Feedback control.

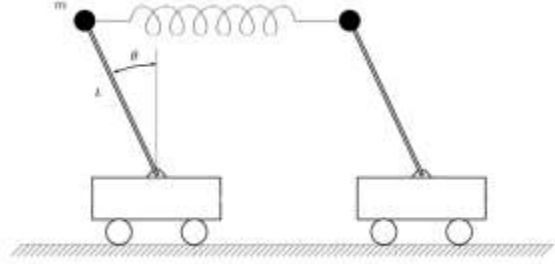

Fig. 1. A concise framework on the decentralized State Feedback control.

In this simulation, the masses of two pendulums chosen as $m_1 = 2kg$ and $m_2 = 2.5kg$; the moments of inertia are $J_1 = 2kg.m^2$ and $J_2 = 2.5kg.m^2$; the constant of the connecting torsional spring is $k = 8N/m$; the length of the pendulum is $r = 1m$; the gravity constant is $g = 9.8m/s^2$. We choose two local models, i.e., by linearizing the interconnected pendulum around the origin and $x_{i1} = (\pm 88°, 0)$, respectively, each pendulum can be represented by the following IT2 T-S fuzzy model with two fuzzy rules.

Rule $l$: IF $A_{il}x_i(t)$ is $F_i^l$ THEN

$$\dot{x}_i(t) = \sum_{l=1}^{r}\sum_{j=1}^{r}\widetilde{w}_{il}\widetilde{m}_{ij}\left\{(A_{il} + B_{il}G_{ij})x_i(t) + D_{1il}\omega_i(t) + \sum_{\substack{k=1 \\ k\neq i}}^{N}\bar{A}_{ik}x_k(t)\right\}$$

$$z_i(t) = \sum_{l=1}^{p}\sum_{j=1}^{c}\widetilde{w}_{il}\widetilde{m}_{ij}\{C_{il}x_i(t) + D_{2il}\omega_i(t)\} \quad (56)$$

where

$$A_{11} = \begin{bmatrix}0 & 1\\ 8.81 & 0\end{bmatrix}, A_{12} = \begin{bmatrix}0 & 1\\ 5.38 & 0\end{bmatrix}, \bar{A}_{12} = \begin{bmatrix}0\\ 0.25\end{bmatrix}, B_{1l} = \begin{bmatrix}0\\ 0.5\end{bmatrix}, \quad D_{1l} = \begin{bmatrix}0\\ 0.5\end{bmatrix}, \quad C_{1l} = [1\ 1] \quad (57)$$

for the first subsystem, and

$$A_{21} = \begin{bmatrix}0 & 1\\ 9.01 & 0\end{bmatrix}, A_{22} = \begin{bmatrix}0 & 1\\ 5.58 & 0\end{bmatrix}, \bar{A}_{21} = \begin{bmatrix}0\\ 0.20\end{bmatrix}, B_{2l} = \begin{bmatrix}0\\ 0.5\end{bmatrix}, \quad D_{2l} = \begin{bmatrix}0\\ 0.5\end{bmatrix}, \quad C_{2l} = [1\ 1] \quad (58)$$

for the second subsystem.

The normalized type-2 membership functions shown in Fig. 2 where $r_i = 88°$.

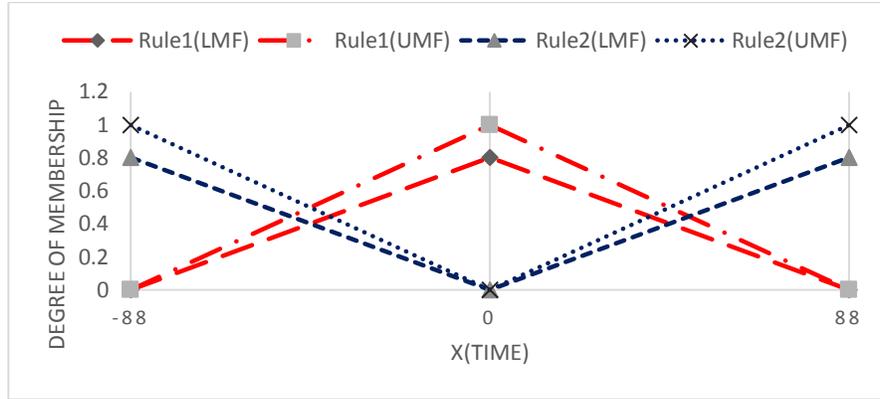

Fig. 2. IT2 Membership function in Example.

Given the initial conditions $x_1(0) = [1.2,0]^T$, $x_2(0) = [0.8,0]^T$ Fig. 3 shows that the double-inverted pendulum system is not stable in the open-loop case.

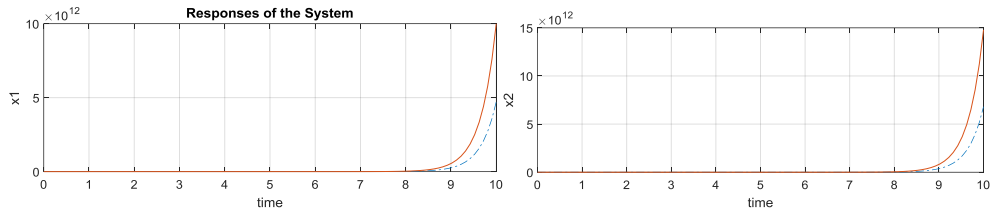

Fig. 3 State responses for open-loop double-inverted pendulums system.

Taken the controller gains solved by Theorem1, Fig. 4 shows the state responses for the closed-loop large-scale system by considering the external disturbances $\omega_1(t) = 0.8e^{-0.2t}\sin(0.2t)$ and $\omega_2(t) = 0.6e^{-0.2t}\sin(0.2t)$ it can be seen that minimum of $H\infty$ disturbance attenuation level $\gamma_{min} = 0.333$ and the desired controller gains obtained in Table 1.

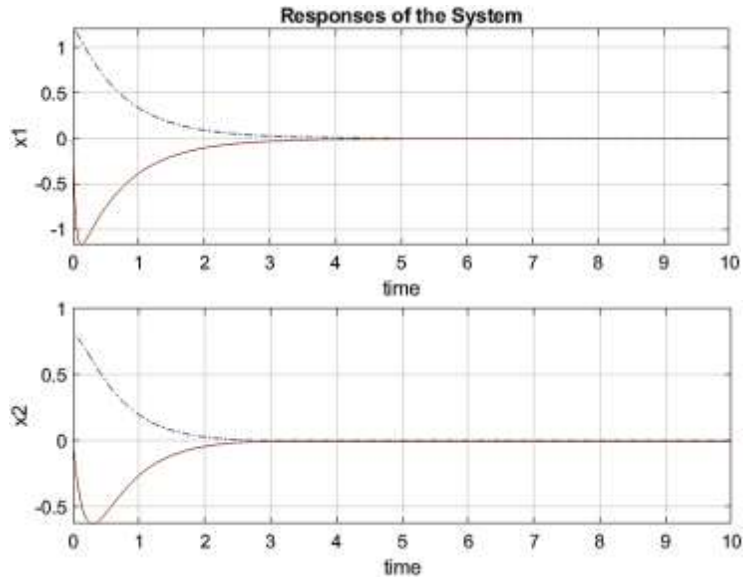

Fig. 4. State responses for closed-loop double-inverted pendulums system based on Theorem1.

Table 1

| $[G_{11}\ G_{12}\ G_{21}\ G_{22}]$ | $\gamma$ |
|---|---|
| $\begin{bmatrix} -34.3381 & -60.0235 & -174.0191 & -485.0611 \\ -16.2743 & -31.7905 & -93.5045 & -268.4191 \end{bmatrix}$ | 0.333 |

## V. Conclusion

In this paper, the robust decentralized state feedback $H_\infty$ type-2 fuzzy controller design has been investigated for continuous-time large-scale type-2 Takagi-Sugeno fuzzy systems. Through some linear matrix inequality techniques, it has been shown that the state fuzzy controller gain can be calculated by solving a set of LMIs. Then the resulting closed-loop fuzzy control system is asymptotically stable under extended dissipativity performance indexes. Uncertainty in the model of large-scale systems is the result of the use of the type-1 fuzzy Takagi-Sugeno model. Therefore, in this paper, the type-2 fuzzy model is used to cover modeling uncertainty for large scale systems. We also stabilize the large-scale system by using the type-2 fuzzy state feedback controller model with imperfect premise membership functions. The advantage of using membership function information in sustainability analysis is to reduce the conservatism of the obtained conditions. Then, in order to reduce the effect of external perturbations on the large-scale system, we applied the robustness control criterion name as extended dissipativity performance indexes to stability analysis, which was able to guarantee the $H_\infty$ criterion, the $L_2 - L_\infty$, passive, and dissipativity performances. Finally, a numerical example of a double-inverted pendulum system has been concerned to verify the effectiveness of the developed methods. The result of this simulation is to improve the control characteristics and make the conditions relax, as well as more complete coverage of the uncertainties in the system. An interesting problem for future research is to deal with the robust decentralized static output feedback $H_\infty$ type-2 fuzzy control design for large-scale systems.

## VI. Appendix

**Assumption 1.** ([22]) Let $\phi, \psi_1, \psi_2$ and $\psi_3$ be matrices such that the following conditions hold:

(1) $\phi = \phi^T, \psi_1 = \psi_1^T$ and $\psi_3 = \psi_3^T$;
(2) $\phi \geq 0$ and $\psi_1 \leq 0$;
(3) $\|D_{2_i}\| . \|\phi\| = 0$;
(4) $(\|\psi_1\| + \|\psi_2\|) . \|\phi\| = 0$;
(5) $D_{2_i}^T \psi_1 D_{2_i} + D_{2_i}^T \psi_2 + \psi_2^T D_{2_i} + \psi_3 > 0$.

**Definition 1.** ([22]) For given matrices $\phi, \psi_1, \psi_2$ and $\psi_3$ satisfying Assumption 1, system (5) is said to be extended dissipative if there exists a scalar $\rho$ such that the following inequality holds for any $t > 0$ and all $\omega(t) \in \mathcal{L}_2[0, \infty)$:

$$\int_0^t J(s)\, dt - z^T(t)\phi z(t) \geq \rho, \tag{59}$$

where $J(t) = z^T(t)\psi_1 z(t) + 2z^T(t)\psi_2 w(t) + w^T(t)\psi_3 w(t)$.

It can be seen from Definition 1 that the following performance indexes hold.

(1) Choosing $\phi = 0, \psi_1 = -I, \psi_2 = 0, \psi_3 = \gamma^2 I$ and $\rho = 0$ the inequality (59) reduces to the $H\infty$ performance [13].
(2) Let $\phi = I, \psi_1 = 0, \psi_2 = 0, \psi_3 = \gamma^2 I$ and $\rho = 0$ the inequality (59) becomes the $L_2 - L_\infty$ (energy-to-peak) performance [14].
(3) If the dimension of output $z(t)$ is the same as that of disturbance $w(t)$, then the inequality (59) with $\phi = 0, \psi_1 = 0, \psi_2 = I, \psi_3 = \gamma I$ and $\rho = 0$ becomes the passivity performance [15].
(4) Let $\phi = 0, \psi_1 = -\epsilon I, \psi_2 = I, \psi_3 = -\sigma I$ with $\epsilon > 0$ and $\sigma > 0$, inequality (59) becomes the very-strict passivity performance [16].
(5) Let $\phi = 0, \psi_1 = Q, \psi_2 = S, \psi_3 = R - \alpha I$ and $\rho = 0$, inequality (59) reduces to the strict $(Q, S, R)$-dissipativity [17].

**Lemma 1.** [23] (Jensen's inequality) For any constant positive semidefinite symmetric matrix $W \in \mathbf{R}^{n \times n}$, $W^T = W \geq 0$ two positive integers $d_2$ and $d_1$ satisfy $d_2 \geq d_1 \geq 1$ then the following inequality holds

$$\left(\sum_{k=d_1}^{d_2} x(k)\right)^T W \left(\sum_{k=d_1}^{d_2} x(k)\right) \leq \bar{d} \sum_{k=d_1}^{d_2} x^T(k) W x(k)$$

where $\bar{d} = d_2 - d_1 + 1$.

**Lemma 2.** for given matrices $\bar{x} \in R^n$, $\bar{y}$ and scaler $\kappa > 0$ we have

$$2\bar{x}^T \bar{y} \leq \kappa^{-1} \bar{x}^T \bar{x} + \kappa \bar{y}^T \bar{y}$$